\documentstyle[11pt,dina4]{article}
\begin{document}
cond-mat/9410074
\begin{center}
{\LARGE \bf \em Ab initio \em simulations of liquid NaSn alloys:\\
Zintl anions and network formation} \\ ~\\ ~\\
M. Sch\"one, R. Kaschner, G. Seifert \\ \em Institut f\"ur Theoretische
Physik \\
Technische Universit\"at Dresden \\
D-01062 Dresden \em \\
\end{center}

\begin{abstract}

Using the Car-Parrinello technique,
\em ab initio \em molecular dynamics simulations are performed
for liquid NaSn alloys in five
different compositions (20, 40, 50, 57 and 80 \% sodium).
The obtained structure factors agree well with
the data from neutron scattering
experiments. The measured prepeak in the structure factor is reproduced
qualitatively for most compositions. The calculated and
measured positions of all peaks show the same trend as function
of the composition.\\
The dynamic simulations also yield information about the formation
and stability of Sn$_4$ clusters (Zintl anions) in the liquid.
In our simulations of compositions with 50 and 57~\% sodium we observe the
formation of networks of tin atoms.
Thus, isolated tin clusters are not stable in such
liquids.
For the composition with 20 \% tin only isolated atoms or
dimers of tin appear, ``octet compounds'' of one Sn atom surrounded
by 4 Na atoms are not observed.

\end{abstract}
\vspace{2cm}
PACS numbers: 61.20.Ja, 61.25.Mv, 71.25.Lf
\newpage

\section{Introduction}

A large amount of work has been done for the past decades on so-called
Zintl-systems; for a review see
 \cite{Review}. In general, these systems are binary alloys which contain
the (so-called) Zintl-ions \cite{Zintl}. These (an)ions are characterized
as isoelectronic to typical neutral
molecular configurations as, for example, the P$_4$ molecule.

The alloys of alkali metals with post-transition group IV elements
are typical examples of Zintl systems: here the atoms of the
group IV elements form anionic Zintl
clusters (especially, tetramer anions isoelectronic to P$_4$). These alloys
played a central role in the development of models for compound formation
and therefore of the concept of polyanion formation (anion clustering)
in liquid anionic alloys.

In this paper we consider the alloys of sodium and tin for which
the first indication of anion clustering in those
systems was found \cite{vdMarel, Geertsma}. The Na-Sn alloys were
investigated experimentally by Alblas et al. \cite{Alblas};
the structures of their solid phases were determined experimentally by
M\"uller and Volk \cite{MV1,MV2}. Theoretical examinations of these
alloys were mostly done by means of simple models \cite{Alblas}.
A first \em ab initio \em simulation with the method of Car and
Parrinello (CP) \cite{CP}
of the equiatomic NaSn liquid was published in \cite{Seifert}.
Here we extend these CP calculations to different compositions
of liquid Na-Sn alloys and discuss their structures, pair correlation
functions and structure factors systematically. In particular, we
investigate the stability of the Zintl anions Sn$_4^{4-}$
and other clusters in the liquids
as a function of the sodium (or tin) concentration. In this paper we
consider
alloys of compositions with 20, 40, 50 (again), 57, and 80 per cent of
sodium.
For these concentrations experimental (neutron scattering) data are
available \cite{Alblas}.

Similar work has been done with \em ab initio \em molecular dynamics
for the alloys K-Si \cite{Galli}, Li-Si \cite{LiSi} and Cs-Pb \cite{CsPb}
and with empirical molecular dynamics for alkali-Pb alloys \cite{EmpMD}.

The method and computational details are shortly summarized
in the next Section. In Sec. 3.1 the results for the structure factors
are presented and compared with the experimental results.
The short-range order, i.e. the cluster stabilities and coordination
numbers, are discussed in Sec. 3.2. Finally a summary is given.

\section{Method and Computational details}

\subsection{Method}
For our simulations we used the Car-Parrinello method \cite{CP}, which was
found to be an effective tool for performing
\em ab initio \em molecular dynamics; for
details of this scheme see \cite{CP1}.
We applied the MOTECC-90 computer code \cite{Hohl}. The pseudopotentials
of Bachelet et al. \cite{BHS} were used, and
the exchange and correlation (XC) energy was treated within the
local density approximation (LDA).
The plane wave basis was taken up to a
cutoff energy of \mbox{6 $Ryd$}, which we found to be sufficient
(see below). For the Brillouin zone sampling in the k-space summation
for the calculation of the electronic density
we only used the $\Gamma$ point (like in \cite{Seifert}).

To determine the cutoff energy and to estimate the error due to
supercell, pseudopotential and LDA effects we calculated the following
quantities:

(i) The equilibrium bond
lengths of the dimers Na$_2$, Sn$_2$ and NaSn with 3 different cutoff
energies
using a large supercell, to avoid spurious interactions between the dimers.
Furthermore, these bond lengths are calculated with an LCAO
program \cite{LCAO} (where the dimers are considered without supercell
effects)
with (a) LDA and (b) LDA plus gradient corrections (GC).

The results - including the
available experimental data - are given in Table 1. The bond lengths,
obtained
by the plane-wave pseudopotential (PWPP) scheme, applied in this work,
agree within about 5 to 7\% with the experimental and LCAO values
and are, in general, slightly too small.
The effect due to the gradient corrections is small compared to this
deviation ($\le$ 3\%).
It is obvious from Tab. 1 that a cutoff energy of 6 $Ryd$ is sufficient.

(ii) The equilibrium lattice constants of the supercell of the Na and Sn
bulk, respectively, for different cutoff energies.
The lattice constants obtained show a similar trend as the
dimer bond lengths. They converge in a similar manner with
the cutoff energy as the dimer bond lengths and are
about 5 to 10 \%  smaller than the experimental values.

\subsection{Simulation Procedure}

As starting configuration for the simulations we took a cubic supercell
with 64 atoms. For the composition with 50 \% Na we used the geometry of
the solid phase of $\beta$-NaSn
as described by M\"uller and Volk \cite{MV1} rescaled to a cubic
cell with a length of $a=23.4 \, a.u.$, keeping the
same volume as the original cell. Therefore, in the case of the equiatomic
composition we used the experimentally known density
of the solid \cite{MV1}; see also \cite{Seifert}. This geometry contains
Sn tetrahedra which form Zintl tetramer anions.

Other compositions were achieved by replacing in the same cell with the same
atomic positions some atoms from Na to Sn or vice versa. Specifically,
we used the following cell configurations: Na$_{52}$Sn$_{12}$ for 80\% Na,
Na$_{36}$Sn$_{28}$ for 57\% Na, Na$_{32}$Sn$_{32}$ for 50\% Na,
Na$_{28}$Sn$_{36}$ for 40\% Na, and Na$_{13}$Sn$_{51}$ for 20\% Na.
As the the effective atomic radii of Na and Sn are very similar
the cell size was kept equal for all compositions.

Having assigned the initial positions, the geometry of the starting
configuration was relaxed with a steepest-descent technique for the
electronic and ionic systems.
After this, the system was equilibrated in order to obtain the
desired average temperature which was chosen to be the temperature for
which
the neutron scattering experiments of \cite{Alblas} have been performed:
Firstly the system was heated to a finite temperature by scaling the
velocities of the nuclei.
Then the Car-Parrinello (CP) dynamics was started for some 500 to 1000
steps of molecular dynamics.
After that the velocities were scaled again, and the
``heating-equilibration'' process was repeated until the averaged
temperature over the equilibration CP run
was near (within $\pm$ 50 K) the desired value.

During this process the total (kinetic and potential) energy $E_{cons}$ of
the nuclei had to be monitored.
The parameters - the time step $\delta t$ and the fictitious mass of the
orbitals $\mu$ - were optimized on the one hand to get a simulation with a
sufficiently  constant energy $E_{cons}$ - basically this means: with small
(fictitious) ``orbital kinetic energies'' - and on the other hand to get a
simulation time large enough to extract physical quantities of a liquid.
Time steps $\delta t$ of 5 to 10 atu with a $\mu$ in the range of 300 to
500 a.u. proved to be suitable ($1 atu = 2.4 \cdot 10^{-17}s$).

After that the data for the forthcoming analysis could be collected from a
``production run'' of at least  10000 steps of CP dynamics.
Hence, the total simulation time was about 2 picoseconds. Although this
time is too small to simulate a real flow of the liquid, it is sufficient
to have an evolution of the system over a time long enough to
describe the fluctuations of the interatomic distances. The CPU time per
simulation (per production run) was approximately 7 days on an IBM-RS6000
workstation.

As check of the ``liquidness'' of the system the time dependence of the
mean
square displacement of the atoms from their starting positions was
monitored.
In liquids, this property increases nearly proportionally with the
simulation time.

Each ``production run'' yielded a phase-space trajectory of the system.
{}From the atomic positions of this trajectory the pair
correlation functions and the structure factors were obtained as described
for
example in \cite{Alblas}. Furthermore, the coordination numbers and the
bond
angle distributions between the atoms could be investigated via a
nearest-neighbor
analysis; these quantities give information for example about the stability
of clusters in the
liquid. These results are presented and discussed in the following.

\section{Results and Discussion}

\subsection{Structure Factors}

The calculated structure factors for the five considered compositions are
plotted in \mbox{Fig. 1} together with the experimental ones. The
experimental
curves were obtained for temperatures of 25 K above the liquidus
temperature.
These temperatures and the final averaged temperatures of the ``production
runs''
corresponding to our curves are given in Table 2.

The general agreement between calculations and experiment is good for all
cases.
The trends of the peak positions as function of the concentration are
reproduced rather well - see Fig.~2.
However, the calculated peak positions (except for the prepeak) are
slightly shifted towards larger k-values compared to the experimental ones,
see
Fig. 2. Such an effect was also found in the systems Li-Si \cite{LiSi} and
Cs-Pb \cite{CsPb}.
This shift can be explained by the occurence of too small distances in the
real space because these distances correspond with too large
distances in the reciprocal lattice.
As can be seen from Table 1, the interatomic distances (bond
lengths) obtained from PWPP calculations are too small. Hence, the shift in
the structure factor could be due to pseudopotential effects.
Another possible explanation could be that the density of the system used
for our simulations is too large compared to the experiments.

For each system the simulation yields a prepeak. In general, the prepeak
is connected to a superstructure in the liquid. Alblas et al. suggested
that the prepeak
corresponds to an ordering of Sn clusters \cite{Alblas}. However, as
already shown in \cite{Seifert}, the existence of a prepeak is not
necessarily
connected with the ``survival'' of such clusters (tetrahedra). For further
discussion of these Zintl clusters see the next subsection.

The calculated positions of the prepeaks show the right trend, see Fig. 2.
However, for small Na concentrations (20 and 40 \%) no prepeak is measured;
furthermore, the heights of the prepeaks have the wrong trend.
The reason for these differences could be the statistical
noise in our simulated structure factors; a too small supercell,
i.e. a too broad mesh of k-points for small k; or a ``superstructure
effect'' of the periodicity imposed by the supercell treatment of the
system, which could cause a prepeak for all
compositions independently from the experiment. On the other hand,
the measured structure factors show for the small Na concentrations a
rather strong noise for small k-values;
therefore a prepeak for these two compositions might be possible, too.

Furthermore, we investigated the temperature dependence of the structure
factors. It was found for each composition that the structure factor
changes only within our numerical accuracy (i.e. within the
statistical fluctuations) for  temperature differences
up to 100 K. Such small changes with the temperature agree with the
measurements \cite{Alblas}.

\subsection{Short-Range Order}

The structure factors obtained from neutron scattering experiments  as
performed in \cite{Alblas} yield no direct information about the geometric
structure of the substances.
In MD simulations one gets the complete structure information about the
system and so it is possible to analyze for example the short-range order
which could be helpful for the understanding of certain physical
properties.

In this paper we mainly focus on the Sn-Sn correlations. Figures 3a-c give
snapshots of typical system configurations with 20, 50 and 80 \% sodium,
respectively, taken from our simulations. The bonds between Sn atoms
with a distance less than 6 a.u. are shown.

In systems with excess tin (20 and 40 \% sodium) the Sn atoms form
``dynamic''
networks characterized by strong fluctuations in the bondings. We observed
only a few atoms with more than fourfold coordination and about as much
with two- as with threefold coordination. This and the bond angle
distributions indicate that the Zintl anions (Sn$_4^{4-}$),
with which our simulations
started, now form larger networks of tin. There is no evidence for the
existence of isolated Zintl anions in these systems.
Fig. 3a gives an impression on how complex the networks are in the case of
the configuration with 20 \% sodium.

The analysis of the data taken from the simulation with 57 \% sodium
support the results  for the equiatomic composition already published in
\cite{Seifert}: The Sn atoms form larger networks than in the case
discussed above.
The Sn$_4^{4-}$ Zintl anions which exist in the solid phase \cite{MV1} do
practically not
survive in the liquid. Only a few Sn$_4$ tetrahedra remain, but they are
parts of a network and not isolated; see Fig. 3b.
This trend to network formation is again supported by the bond angle
distribution analysis, see \cite{Seifert}.
Similar results were found for the alloys K-Si \cite{Galli}, Li-Si
\cite{LiSi}, and Cs-Pb \cite{CsPb}. Their Si- and Pb-
tetrahdra, respectively, tend to build blocks (networks) in the liquid.

When we reduced the share of tin in the system (80 \% Na) neither networks
nor Zintl Anions could be observed. Instead we found that the tin atoms
preferably stay isolated or form dimers. There is no remarkable trend for
a dimer seperation, but some evidence for the formation of new dimers
(``dumb-bells'') during our simulations (see Fig. 3c).
So these dimers seem to be relatively stable with an oscillating bond
length with an average of about 5.5 a.u.

For this composition we also investigated the existence and stability of
the
so-called ``octet compounds'' (Na$_4$Sn - clusters) which were suggested by
Alblas et al. (\cite{Alblas}). From our data we could not confirm the
existence of such stable complexes.

More detailed discussions on the pair correlation functions, bond angle
distributions and coordination numbers and especially results on the
electronic structure will be given in a forthcoming  paper \cite{Paper}.

\section{Summary, Outlook}

In this paper five different compositions of liquid Na-Sn alloys were
investigated
by means of \em ab initio \em (Car-Parrinello) simulation. The obtained
structure factors agreee well with the experimental data. The peak
positions show the correct trend as functions of the concentration.
The systematic shift can be explained as (small)
pseudopotential effect.
In particular, the measured prepeak could be reproduced. Deviations between
simulation and experiment concerning the height or the disappearance
of the prepeak could be due to
a noise in the experiment or to statistical fluctuations in the simulation.
The structure factors show a very small temperature dependence  - within
the
statistical noise of our simulations which is also in agreement with the
experiment.

We find that Sn$_4$ clusters are not stable but form networks
in the liquid with 50 and \mbox{57 \%} Na. In the case of 80 \% Na no
stable Na$_4$Sn clusters are observed - which disagrees with suggestions of
Alblas \cite{Alblas}:
Their picture of isolated Sn$_4$ tetrahedra and Na$_4$Sn octet compounds,
respectively, does not seem to be an appropiate description of the atomic
structure of liquid NaSn alloys.

Our theoretical method has the following limitations: (i) the (finite)
supercell,
which leads to a broad mesh of the k points for small k, (ii) the finite
simulation time which causes a statistical noise of the results, (iii) the
limited accuracy due to the
pseudopotential  - which we used without the so-called
``non-linear core-corrections'' \cite{corecorr} to have a smaller cutoff
energy to economize computer time. To overcome these problems one could use
a softer pseudopotential, see e.g. \cite{PPnew}, or one looks for
approximate
methods which allow the simulation of larger times and larger supercells.

On the other hand, our method has the advantage that the electronic
structure is considered directly, without empirical parameters (\em ab
initio \em method). Furthermore, the resulting trajectories of the
\em ab initio \em simulations
with the corresponding electronic structure (wavefunctions, eigenvalues)
allow
to calculate other quantities of physical interest, as, for example, the
resistivity,
charge distribution, specific heat, susceptibilities and the Knight shift.
For some of these quantities experimental data are
available. Especially the resistivity (or conductivity)
is known to show a strong dependence on changes in composition.

As next steps, we are going to
extract these quantities from our data.
Furthermore, we may extend our investigations to other binary (Zintl)
alloys.\\

Acknowledgement: The authors thank D. Hohl, Institut f\"ur
Festk\"orperforschung, J\"ulich, for providing us with his Cornell version
of the CP code based on the MOTECC-90 program collection by IBM
and are grateful to G. Pastore, ICTP Trieste, for helpful
discussions and support concerning the computer programs.
This work was supported by the Deutsche Forschungsgemeinschaft (DFG).

\newpage

\vspace*{2cm}
\begin{center}
\begin{tabular}{|c|c|c|c|c|c|c|}
\hline
& \multicolumn{3}{|c|}{{\large PWPP - LDA}} & \multicolumn{2}{|c|}{{\large
 LCAO}} & {\large Exp.} \\
\hline
                 &  {\large ~6 Ryd}   & {\large ~8 Ryd} & {\large 10 Ryd}
		 & {\large LDA}  & {\large GC}    & \\
\hline
{ Na${_2}$}     &  { 5.484}  & { 5.483} & { 5.482}  & { 5.67} & { 5.84}
& { 5.82} \\
\hline
{ Sn${_2}$}     &  { 5.014}  & { 5.007} & { 5.030}  & { 5.29} & { 5.37}
& { 5.30} \\
\hline
{ NaSn}       &  { 5.430}  & { 5.439} & { 5.440}  & { 5.69} & { 5.83}
&  \\
\hline
\end{tabular}
\end{center}
\vspace*{1cm}
Table 1: Equilibrium bond length (in a.u.) for the dimers
Na$_2$, Sn$_2$ and NaSn calculated with several cutoff energies by the
PWPP method  and the LCAO method (explanation see text) compared to
experimental values from [18].
\vspace*{2cm}
\begin{center}
\begin{tabular}{|c||c|c|c|c|}
\hline
{\large System} & {\large atomic \% Na} & {\large $\bar T_{MD}$
[K]} & {\large $T_{exp}$ [K]} & {\large $T_{liq}$ [K]} \\
\hline
\hline
Na$_{13}$Sn$_{51}$ &   20.3   &   726   &   623   &   580   \\
\hline
Na$_{26}$Sn$_{38}$ & { 40.6 } & { 818 } & { 813 } & { 783 } \\
\hline
Na$_{32}$Sn$_{32}$ & { 50.0 } & { 921 } & { 873 } & { 851 } \\
\hline
Na$_{36}$Sn$_{28}$ & { 56.3 } & { 796 } & { 778 } & { 752 } \\
\hline
Na$_{52}$Sn$_{12}$ & { 81.3 } & { 748 } & { 773 } & { 681 } \\
\hline
\end{tabular}
\end{center}
\vspace*{1cm}
Table 2: Systems and temperatures discussed in this paper. \\
For each system the averaged temperature $\bar T_{MD}$ of the simulation
is compared to the temperature $T_{exp}$ for which neutron scattering
experiments in [5] have been performed.\\
In addition the liquidus temperature $T_{liq}$ is given.
\newpage
Figure 1: Static structure factors obtained from  CP - simulations compared
to those from neutron scattering experiments ([5]) for different NaSn
alloys.
Given is the atomic percentage for each composition. For details on the
systems see Table 2.

\vspace*{2cm}
Figure 2: Peak positions in the static structure factors shown in Fig. 1
plotted as function of the sodium concentration. q$_p$: prepeak, q$_1$:
main peak (maximum), q$_2$: next minimum, q$_3$: next maximum. \\
The lines indicate the trend of the peak positions.
Dashed lines refer to experimental data [5], solid lines refer to
CP - simulations.

\vspace*{2cm}
Figure 3: Snapshots of system configurations with 20 (a), 50 (b) and 80
(c) \% sodium. The Na atoms are plotted as smaller and brighter.
The Sn-Sn bonds within the cut-off distance of 6 a.u. are shown.

\end{document}